\documentclass[12pt]{JHEP3}
\usepackage{amsmath,amsfonts,amssymb,amsthm,amstext,bbm, amscd,eucal}
\usepackage{epsfig}
\usepackage{graphicx}
\usepackage{psfrag}
\usepackage{epstopdf}

\def\nn{\nonumber}

\newcommand{\mpl}{M_{\rm pl}}

\def\AS{{\alpha_k^S}}
\def\A{{\alpha_k}}
 \def\B{{\beta_k}}
 \def\AT{{\alpha_k^T}}
 \def\BS{{\beta_k^S}}
\def\BT{{\beta_k^T}}
\def\chit{\chi_{{}_T}}
\def\chis{\chi_{{}_S}}
\def\phit{\varphi_{{}_T}}
\def\phis{\varphi_{{}_S}}

 \newcommand{\be}{\begin{equation}}
\newcommand{\ee}{\end{equation}}
\newcommand{\beqa}{\begin{eqnarray}}
\newcommand{\eeqa}{\end{eqnarray}}

\title{\normalsize{\Large{Reconciliation of High Energy Scale Models of Inflation with Planck}}}

\author{Amjad Ashoorioon \footnote{a.ashoorioon@lancaster.ac.uk}\\
{Physics Department, Lancaster University, Lancaster, LA1 4YB, United
Kingdom}}

\author{Konstantinos Dimopoulos \footnote{konst.dimopoulos@lancaster.ac.uk}\\
{Physics Department, Lancaster University, Lancaster, LA1 4YB, United
Kingdom}}

\author{ M. M. Sheikh-Jabbari  \footnote{jabbari@theory.ipm.ac.ir}\\
 School of Physics, Institute for Research in Fundamental Sciences (IPM), Tehran, Iran}

\author{Gary Shiu \footnote{ shiu@physics.wisc.edu }\\
{Department of Physics, University of Wisconsin, Madison, WI 53706 and\\
Center for Fundamental Physics and Institute for Advanced Study,
Hong Kong University of Science and Technology, Hong Kong}}

\abstract{The inflationary cosmology paradigm is very successful in explaining the CMB anisotropy to the percent level.
Besides the dependence on the inflationary model, the power spectra, spectral tilt and  non-Gaussianity of the CMB temperature fluctuations also depend on the initial state of inflation. Here, we examine to what extent these observables are affected
by our ignorance in the initial condition for inflationary perturbations, due to unknown
new physics at a high scale $M$. For initial states that satisfy constraints from backreaction, we find that the amplitude of the power spectra could still be significantly altered, while the modification in bispectrum  remains small. For such initial states, $M$ has an upper bound of a few tens of $H$, with $H$  being the Hubble parameter during inflation. We show that for $M\sim  20 H$,  such
initial states  always (substantially) suppress the tensor to scalar ratio. In particular we show that such a choice of initial conditions can  satisfactorily reconcile the simple $\frac{1}{2}m^2 \phi^2$ chaotic model with the Planck data \cite{Planck-data}
}


\preprint{\today }

\begin{document}

\parskip 6 pt
\lineskip 2pt

\section{Introduction}

\begin{figure}[ht]
\centerline{\includegraphics[angle=0,scale=1.1]{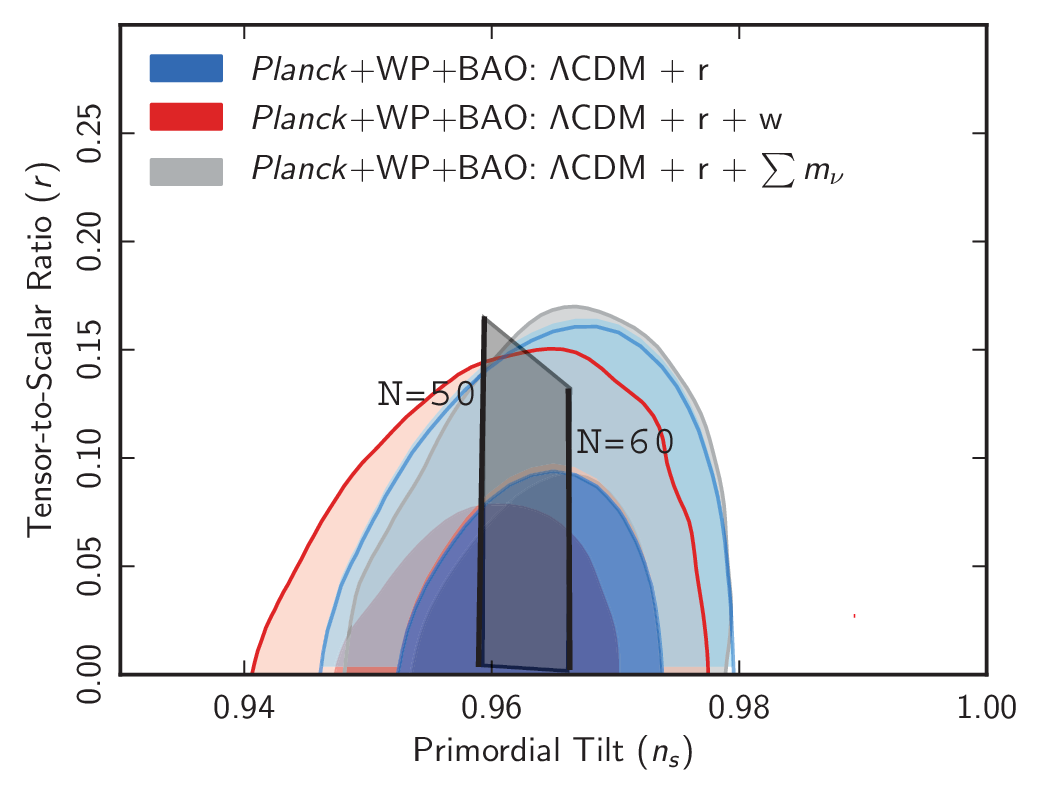} }
\caption{$n_s$ vs. $r$ $1\sigma$ and $2\sigma$ contours, Planck results \cite{Planck-data}. The grey shaded trapezoid region shows the acceptable region for the $m^2\phi^2$ model. The upper oblique line corresponds to the BD vacuum and the Left and Right sides correspond to $N_e=50$ and $N_e=60$ respectively.}\label{ns-r-gf}
\end{figure}

The increasingly precise cosmic microwave background (CMB)  measurements \cite{Planck-data}, in combination with
 other cosmological data, have ushered us into a precision early Universe cosmology era: The power spectrum of CMB temperature fluctuations is measured to have an amplitude of order $10^{-5}$. The spectrum is almost flat, with a few-percent tilt toward larger scales (i.e., red spectrum)
and is almost Gaussian. Moreover, the B-mode polarization in the CMB fluctuations, which is an indicative of primordial gravity waves, has not been observed with $10\%$ accuracy or better. A part of this cosmic data, which we will discuss further below, is summarized in Fig. \ref{ns-r-gf}.

Inflation is by far the leading
 paradigm in explaining the CMB temperature anisotropy. This paradigm,
that the early Universe has gone through a period of accelerated expansion,
can be realized in concrete models formulated within effective (quantum) field theory coupled to gravity.
 Inflation thus provides an interesting link between short distance physics and cosmological observations.  In the inflationary picture,
  the CMB fluctuations can be  traced back to the quantum fluctuations  during inflation
 \cite{Mukhanov}:
 quantum effects are stretched to cosmological size due to the quasi-exponential expansion of inflation.
As the pattern of quantum fluctuations depends on the microphysics of
inflation, precision data from the CMB and other cosmological measurements
thus allow us to constrain or rule out inflationary models.

However, predictions of inflationary models for the CMB temperature anisotropy
depend not only on
details of the model itself,
but also
on the initial state of the quantum fluctuations.
 It is usually assumed that these fluctuations start in the Bunch-Davis (BD) vacuum, as
 they are expected to
be in a minimum energy state
 when they are produced inside the horizon  of an inflationary background. However,
 various well-motivated early universe physics,
 e.g.
 effects of high energy cutoff
 \cite{Initial-data-literature}
  and multi-field dynamics \cite{Shiu:2011qw}
can place
these
 fluctuations
 in an excited state, i.e., a non-BD initial state.

In this paper,  we discuss how a non-BD initial state in a consistent theoretical setup can crucially affect the implications of cosmic data for inflationary models. In particular, we show that the simplest inflationary model, the quadratic chaotic inflation model \cite{Linde:1983gd} in BD vacuum, which is disfavored by the current Planck mission data \cite{Planck-data}, can still be a viable model of inflation with a very good agreement with the data if the fluctuation start in a non-BD state\footnote{One way to reconcile these models with the data is to consider gravity as a field which is inherently classical. If such, there will be no quantum tensor seeds which can be stretched to superhorizon scales by the expansion of the universe \cite{Ashoorioon:2012kh}. The tensor/scalar ratio will be exactly zero in that case.}.

The outline of the paper is as follows. First we review the cosmological perturbation theory with an excited initial state. We discuss how backreaction can constraint the deviations from Bunch-Davies vacuum. In the next section, we show how this is translated to bounds on the scale of new physics with respect to the inflationary Hubble parameter. We analyze the parameter space of initial conditions for scalar and tensor fluctuations and show that in the region that scale of new physics is maximally separated from the inflationary parameter, the effect on the tensor over scalar ratio is always suppressive. Finally, in the last section, we show that the large deviations from Bunch-Davies vacuum does not lead to essential enhancement of non-Gaussianity and if a model with BD initial states is compatible  with the Planck data, it will remain so in the non-BD initial state.

\section{Cosmological perturbations with excited initial states}

An excited initial state can in principle affect the background inflationary dynamics (through the backreaction of the perturbations) as well as
the CMB temperature anisotropy. To this end we recall some results of cosmological
 perturbation theory, whose details  may be found in \cite{Mukhanov}.
Let us consider the simplest single scalar field inflationary model:
\begin{equation}
\mathcal{L}=-\frac{\mpl^2}{2}R-\frac{1}{2}\partial _{\mu }\phi \partial ^{\mu }\phi -V(\phi)
\label{action-scalar-minimal}
\end{equation}%
where $\mpl^{-2}=8\pi G_N$ is the reduced Planck mass. While true in more general cases, we will focus on the quadratic chaotic model, $V(\phi)=\frac12m^2\phi^2$, in most of our analysis.
The metric scalar and tensor perturbations can be parameterized as
\begin{equation}\label{mtrc}
ds^{2}=a^{2}(\tau )\left[ -(1+2\Phi ){d\tau }^{2}+\left( (1-2\Psi ){\delta }%
_{ij}+h_{ij}\right) d{y}^{i}d{y}^{j}\right] ,\nn
\end{equation}%
where $\Phi $ and $\Psi $ are the scalar Bardeen potentials and $h_{ij}$ is a symmetric
divergence-free traceless tensor field, $
h_{i}^{i}=0,\ \partial^ih_{ij}=0$. The inflaton
field is also perturbed about its homogeneous background value
\begin{equation}
\phi (\tau )=\phi_{\rm hom.}(\tau )+\delta \phi .  \label{inflat}
\end{equation}%
where
$\phi_{\rm hom.}(\tau)$ is the homogeneous part of the inflation which satisfies $\delta
\phi \ll \phi_{\rm hom.}(\tau)$ and
the
 perturbed Einstein equations imply $\Phi=\Psi$. The equation of motion for the gauge-invariant scalar perturbations, the Mukhanov-Sasaki variable $u(\tau,y)$,
\begin{equation}\label{u-mukhanov}
u=-z \left( \frac{a^{\prime}}{a}\frac{\delta \phi}{\phi^{\prime}}+\Psi\right), \quad z\equiv \frac{a \phi^{\prime}}{\cal H}, \quad {\cal H}\equiv \frac{a^{\prime}}{a},
\end{equation}
is
\begin{equation}\label{u-eq}
u^{\prime\prime}_k+\left(k^2-\frac{z^{\prime\prime}}{z}\right)u_k=0\,,
\end{equation}
where prime denotes derivative w.r.t. conformal time $\tau$ and $u_k(\tau)$ is the Fourier mode of $u(\tau,y)$.
For a quasi-de-Sitter background
 \begin{equation}\label{background}
 a(\tau)\simeq -\frac{1}{H\tau}
 \end{equation}
where $H$ is the Hubble constant and
 \begin{eqnarray}\label{slow-roll-def}
\epsilon\equiv 1-\frac{{\cal H}^{\prime}}{{\cal H}^2}\ll 1\,,\qquad \eta\equiv\epsilon-\frac{\epsilon^{\prime}}{2{\cal H}\epsilon}\ll 1,
\end{eqnarray}
the most generic solution to \eqref{u-eq} in the leading order in slow-roll parameters $\epsilon, \eta$ is of the form of Bessel functions:
\begin{equation}\label{u-sol-ds}
u_k(\eta)\simeq\frac{\sqrt{\pi|\tau|}}{2}\left[\AS~  H_{3/2}^{(1)}(k|\tau|)+\BS H_{3/2}^{(2)}(k|\tau|)\right].
\end{equation}
$H_{3/2}^{(1)}$ and $H_{3/2}^{(2)}$ are respectively  Hankel functions of the first and second kind. The terms proportional to $\AS$ and $\BS$ respectively behave like the positive and negative frequency modes. These Bogoliubov coefficients
satisfy the Wrosnkian (or canonical normalization) constraint
\begin{equation}\label{Wronskian}
|\AS|^2-|\BS|^2=1.
\end{equation}
The standard BD vacuum corresponds to $\AS=1$ and $\BS=0$. As discussed, one may start with a generic non-BD excited initial state, i.e. a generic $\A, \B$.

In the BD vacuum the energy density and pressure carried away from the inflationary background by the frozen-out perturbations in an e-fold is $\delta\rho_0\sim H^4$, where $H$ is Hubble during inflation. The energy density and pressure of the background, on the other hand, as implied by Friedmann equation is $\rho_0=3\mpl^2 H^2$. To make sure that the backreaction is small, $\delta\rho_0$ should be smaller than the decrease in the background energy density due to expansion in an e-fold $\Delta\rho_0\sim \epsilon \rho_0\sim \epsilon H^2\mpl^2$. That is, $\delta\rho_0\ll \Delta\rho_0\sim \epsilon\rho_0$. This condition is satisfied if $ H^2\ll \mpl^2\epsilon$ which is expected to be satisfied for almost all single field inflationary models, recalling the COBE normalization.

For a generic initial state, however, the energy or pressure density carried by the perturbations is
\begin{equation}\label{delta-rho-free}
\delta\rho_{\text{non-BD}}\sim \frac{1}{a(\tau)^4}\int_{\cal H}^{\infty} \frac{d^3 k}{(2\pi)^3}\ \frac12(|\AS|^2+|\BS|^2-1) k\,,
\end{equation}
and $\delta p_{\text{non-BD}} \sim \delta \rho_{\text{non-BD}}$. In the above we have dropped the factor of $z^{\prime\prime}/z$ as the effects we are studying are mainly coming from sub-Hubble modes. One may see that $\delta\rho_{\text{non-BD}}'\sim \delta p_{\text{non-BD}}'\sim {\cal H} \delta\rho_{\text{non-BD}}$ in the leading slow-roll approximation.
The backreaction will not derail slow-roll inflationary background if
\be\label{background-backreaction}
\delta\rho_{\text{non-BD}}\ll \epsilon\rho_0\,,\quad \delta p_{\text{non-BD}}'\ll \mathcal{H} \eta\epsilon\rho_0\,.
\ee
Note that the normalization condition \eqref{Wronskian}, assuming slow-roll, may be written in terms of $\beta_k$
\be
\int_{\mathcal{H}}^{\infty} \frac{d^3 k}{(2\pi)^3} k |\BS|^2\ll \epsilon\eta H^2\mpl^2\,.
\ee

We now analyze the effects of the non-BD initial state on the  scalar and tensor modes power spectra. The scalar power spectrum is defined as
\begin{equation}
{\mathcal P}_{S}=\frac{k^{3}}{2\pi ^{2}}\left| \frac{u_{k}}{z}\right|^2_{{k/{\cal H}\rightarrow 0}}.
\label{scrpower}
\end{equation}
which for simple chaotic models reduces to
\begin{equation}\label{power-spectrum-scalar}
{\mathcal P}_S={\mathcal P}_{BD}\,\gamma_{{}_S} ,
\ee
where
\be
{\mathcal P}_{BD}=\frac{1}{8\pi^2\epsilon}\left(\frac{H}{\mpl}\right)^2,\qquad \gamma_{{}_S}=|\AS-\BS|^2_{{}_{k={\cal H}}}.
\end{equation}
Note that  the dependence of the power spectrum on $\A,\B$ is different than that of the energy density of the modes \eqref{delta-rho-free}. From \eqref{power-spectrum-scalar} one also sees that the spectral tilt $n_s-1\equiv d\ln P_s/d\ln k$ can in principle be affected by the choice of initial conditions. However, in our analysis here we restrict ourselves to cases where the spectral tilt is not affected.

Likewise, one may repeat a similar analysis for the tensor perturbations $h_{ij}$.
The tensor mode perturbations are also
given by Hankel functions:
\begin{equation}\label{p-sol-ds}
h_k(\tau)\simeq\frac{\sqrt{\pi|\tau|}}{2}\left[\AT~  H_{3/2}^{(1)}(k|\tau|)+\BT~ H_{3/2}^{(2)}(k|\tau|)\right],
\end{equation}
where $h_k(\tau)$ is the Fourier mode of the amplitude of either of the gravity wave polarizations. The tensor power spectrum is then
given by
\begin{equation}\label{pwr-tensor}
 {\mathcal P}_{T}={\mathcal P}_{\rm BD}^T\ \gamma_{{}_T}\,,
\ee
and
\be
{\mathcal P}_{\rm BD}^T=\frac{2}{\pi^2}\left(\frac{H}{\mpl}\right)^2\,,\quad \gamma_{{}_T}=|\AT-\BT|^2_{{}_{k={\cal H}}}\,,
\end{equation}
where $\AT$ and $\BT$ satisfy the normalization condition $|\AT|^2-|\BT|^2=1$.
The tensor-to-scalar ratio is then
\begin{equation}\label{consistency}
r\equiv\frac{{\mathcal P}_T}{{\mathcal P}_S}=16 \gamma\epsilon,\quad \gamma=\frac{\gamma_{{}_T}}{\gamma_{{}_S}}=\left|\frac{\AT-\BT}{\AS-\BS}\right|^2\,.
\end{equation}
Since in our analysis the tensor initial state parameters $\alpha^T, \beta^T$ and those of
the
scalars are taken to be independent, we should in principle also check
the smallness of the backreaction, i.e. \eqref{delta-rho-free} and \eqref{background-backreaction}, for the tensor modes.

The factor $\gamma$  which parameterizes
the
effects of initial states can in principle be bigger or smaller than one.
The
$\A, \B$ parameters  need not be the same  for
the scalar and tensor modes. This can be seen from the fact
that they parameterize excitations in the initial state (e.g. caused by new physics at super-Hubble scale $M$). The new physics, which is assumed to have a description in terms of a generally covariant  effective field theory, can affect scalar and tensor sectors differently. Therefore, $\gamma$ is not  necessarily one. If $\gamma$ is smaller than one, it will help to suppress
$r$ in models with large $H$, like chaotic models. This leads to an ``initial state modified'' Lyth bound \cite{Lyth-bound} ({\it c.f.} \cite{Boubekeur:2012xn}) on the
inflaton field range $\Delta \phi$ during inflation
\be\label{Lyth-modified}
r\lesssim 2.5\times 10^{-3} \left(\frac{\Delta\phi}{\mpl}\right)^2\ \gamma\,,
\ee
and also a modified consistency relation $r=-8\gamma n_T $. The above
indicates that super-Planckian field excursions does not necessarily allow for large $r$, if $\gamma\lesssim 1$.

As a specific example, following \cite{Holman:2007na, Boyanovsky:2006qi}, let us consider a crude model with
\be\label{beta-Gaussian}
\beta_k\propto \beta_0 \exp\left\{{{-k^2/\left[M a(\tau)\right]}^{2}}\right\}
\ee
(or any smooth function in which
$\left|\beta_k\right|^2$ falls off as $k^{-(4+\delta)}$). The above form roughly implies
that the non-BD state kicks in at scales above $M$ which is the scale of new physics  and $\tau_0$ marks the moment the physical momentum of the mode becomes of order of the new physics scale, $k/a(\tau_0)\sim M$. This choice, hence leads to no extra $k$-dependence in power spectra and does not change the spectral tilt. Next, we note that
\begin{equation}\label{p-rho-massless-quanta}
\delta \rho_{\text{non-BD}} \sim \left|\beta_0\right|^2 M^4\,,\quad
\delta p_{\text{non-BD}}'/{\cal H}\sim \left|\beta_0\right|^2  M^4.
\end{equation}
Thus one obtains the following upper bound on $\left|\beta_0\right|$,
\begin{eqnarray}\label{beta-scalar-backreaction}
   \beta_0 \lesssim \sqrt{\epsilon\eta}\frac{H M_{\rm Pl}}{M^2}\sim \epsilon\frac{H M_{\rm Pl}}{M^2}.
\end{eqnarray}
As we will discuss, $\beta_0$ is not necessarily very small. Moreover, the above indicates that the upper bound on the deviation from BD initial state, measured by $\beta_0$, is inversely proportional to the scale of new physics $M$. Hence, larger values of $M$ require smaller $\beta_0$.

\section{Allowed region in non-BD initial states parameter space}

To study this more closely, we note that the energy and the power spectra (and also the bi-spectrum) expressions only depend on relative phase of $\alpha,\ \beta$. Hence, they may be parameterized as
\be\label{parametrization}
\begin{split}
\alpha^S_k =\cosh\chi_{{}_S} e^{i\varphi_{{}_S}}\,&,\quad \beta^S_k =\sinh\chi_{{}_S} e^{-i\varphi_{{}_S}}\,,\\
\alpha^T_k =\cosh\chi_{{}_T} e^{i\varphi_{{}_T}}\,&,\quad \beta^T_k =\sinh\chi_{{}_T} e^{-i\varphi_{{}_T}}\,,
\end{split}
\ee
With this parametrization, $\chi_{{}_S}\simeq\sinh^{-1}\beta_0,\ e^{-2\chi_{{}_S}}\leq \gamma_{{}_S}\leq e^{2\chi_{{}_S}}$, and $e^{-2\chi_{{}_T}}\leq \gamma_{{}_T}\leq e^{2\chi_{{}_T}}$. Using the COBE normalization, assuming $\epsilon\sim \eta\sim 0.01$ we learn that
\be\label{H/Mpl}
\frac{H}{\mpl}=\frac{1}{\sqrt\gamma_{{}_S}} 3.78\times 10^{-5}\,,
\ee
and the backreaction condition \eqref{beta-scalar-backreaction} reads
\be\label{M/H}
\frac{M^2}{H^2}\lesssim 220 \frac{\sqrt\gamma_{{}_S}}{\sinh\chi_{{}_S}}\,.
\ee
A similar analysis can be carried out for the tensor modes, assuming the same form as \eqref{beta-Gaussian} for the $\beta^T$. While one may choose the scale of new physics $M$ to be different for tensor or scalar modes, for simplicity we choose to work with the same $M$ for both modes. The backreaction for tensor modes is small if
\be\label{tensor-backreaction}
\frac{M^2}{H^2}\lesssim 220 \frac{\sqrt\gamma_{{}_S}}{\sinh\chi_{{}_T}}\,.
\ee

Let us first consider a specific, but illuminating case: $\chit=\chis=\chi$, $\phis=\pi/2$ and $\phit=0$. In this case,
the scalar and tensor backreaction conditions reduce to
\be
\frac{M^2}{H^2}\lesssim 440 \frac{1}{1-\sqrt{\gamma}}\,,\qquad \gamma=e^{-4\chi}\leq 1.
\ee
Demanding $\gamma<0.5$ leads to $M\lesssim 39H$. Note also that one can  decrease $\gamma$ to arbitrary small values and that for this case $H=3.8\times 10^{-5}\gamma^{1/4}\mpl$.

For a more general analysis it is convenient to distinguish two different cases:

$\bullet$ \emph{Quasi-BD case:}, $\chi_{{}_S}\ll 1$ and generic $\phis$. In this case $\gamma_{{}_S}\sim 1$ and as physically expected, $M$ can be arbitrarily large and $H$ is very close to its BD value.

$\bullet$ \emph{Typical  or large $\chi_{{}_S}$:} $\chis \gtrsim 1$. For generic values of  $\phis$, $\sqrt{\gamma_{{}_S}}\sim e^{\chis}\sin\phis$ and $\sinh\chis\sim e^{\chis}/2$. We see from \eqref{M/H}  that $M\lesssim 21 H$.  Also from \eqref{H/Mpl} we note that $H$ can be made (much) smaller than $3.78\times 10^{-5}\mpl$ which is its corresponding BD value. Physically, we    need to ensure that $M\gtrsim H$ and hence $\phis\gtrsim 10^{-3}$. The desirable larger values of $M$, e.g. $M\simeq 20H$, is obtained if $\phis\simeq \pi/2$, i.e. when $\alpha$ and $\beta$ have opposite phase. For $M\sim 20H$, \eqref{tensor-backreaction} is satisfied if
\be\label{tensor-backreaction}
2\sinh\chit \lesssim \sqrt{\gamma_{{}_S}} \simeq e^{\chis}\sin\phis.
\ee

{}From \eqref{tensor-backreaction} we learn that $\chit$ can be in quasi-BD, or typical or large $\chit$ regions.
The tensor to scalar ratio suppression factor $\gamma$ is then
\be
\gamma\simeq\left\{\begin{array}{cc}\frac{e^{-2\chis}}{\sin^2\phis},& \qquad {\chit}\ll 1,\\ \,\,\ & \,\,\ \\
 e^{2(\chit-\chis)}\frac{\sin^2\phit}{\sin^2\phis}, & \quad {\chit}\gtrsim 1,\ \text{generic}\ \phit,
\\ \,\,\ & \,\,\ \\
 e^{-2(\chit+\chis)}\frac{1}{\sin^2\phis}, & \quad {\chit}\gtrsim 1,\ \tan\phit\lesssim e^{-2\chit}
.\end{array}\right.
\ee
That is as long as \eqref{M/H} and \eqref{tensor-backreaction} hold, for any value of $\chit$, $\gamma\lesssim 1$.

Notice that regardless of $\varphi$-values, for large $M$, $M\gtrsim H$, suppression of backreaction of scalar and tensor non-BD excitation results in suppression of tensor-to-scalar ratio compared to its BD value, i.e. $\gamma\lesssim 1$. We also note that  the backreaction considerations only impose an upper bound on the the value of $H$: $H/\mpl\lesssim 3.78\times 10^{-5}$ which is its BD value for the quadratic chaotic model.

Next, we note that the shift in the spectral tilt due to the non-BD vacuum state is given by
\be
\delta_{non-BD}(n_s-1)=\frac{\partial\ln \gamma_{{}_S}}{\partial\ln k}\simeq 2\cot\phis \frac{\partial\phis}{\partial\ln k}\,,
\ee
where we used the value of $\gamma_{{}_S}$ in the typical or large $\chis$ range and assumed $\phis$ to have $k$-dependence.
On the other hand, our earlier arguments indicate that we need to consider $\phis$ close to $\pi/2$. Therefore, for smooth $k$-dependence of $\phis$, the shift in the spectral tilt due to non-BD vacuum should be very small. We will henceforth ignore $k$-dependence in the phase $\phis$.

The results of the $\frac{1}{2} m^2\phi^2$ model may be brought to $1\sigma$ contour of the Planck results (Fig. 1), if $\gamma<0.5$. The above discussions show that this cannot happen for the quasi-BD $\chis$ case. However,  \emph{typical $\chi$'s} values can do the job. For example, for $M\simeq 20H$, $\phis=\pi/2$ and generic values of $\phit$, $\gamma\simeq e^{2(\chit-\chis)}\sin^2\phit$ which can  be made of order $0.5$ or lower.  Smaller values of $\phit$ can reduce $r$  to much smaller values. Also in this case, if $\chit<\chis$ the ratio can lowered further.
As depicted in Fig. 1, non-BD initial states can bring this model back to a favorable region with the recent Planck satellite data.

\section{Non-BD initial state and non-Gaussianity}

In view of the Planck results for non-Gaussianity \cite{Planck-data} one should also make sure that possible enhancement of the three point function of curvature perturbation due to non-BD initial conditions is not large. The computations may be carried out following \cite{Maldacena:2002vr}. Similar analysis with non-BD initial state has appeared in  \cite{Chen:2006nt, Holman:2007na, folded-NG, Ashoorioon:2010xg, Agullo:2010ws, squeezed-NG}. Here, we follow the conventions in \cite{Ashoorioon:2010xg}
where it was shown that
 \begin{eqnarray}\label{zeta5}
&&{\cal B}\equiv \langle \zeta_{\overrightarrow{k_1}}\zeta_{\overrightarrow{k_2}}\zeta_{\overrightarrow{k_3}}\rangle=\frac{\pi^3 H^4}{\mpl^2\epsilon k_1^3 k_2^3 k_3^3} \mathcal{A}\\
&&{\cal A} =(\sum_{i<j} k_i^2 k_j^2)\biggl[\frac{(1-\cos(k_t \tau_0))}{k_t}C_1-\frac{\sin(k_t\tau_0)}{k_t}C_2\nonumber\\  &&\ \ \ +C_3\sum_{j=1}^{3}\frac{(1-\cos(\tilde{k}_j\tau_0))}{\tilde{k}_j}-C_4\sum_{j=1}^{3} \frac{\sin{(\tilde{k}_j\tau_0)}}{\tilde{k}_j}\biggr],
 \end{eqnarray}
Here, $\zeta_{\overrightarrow{k_i}}$ are the curvature perturbations and $k_t=k_1+k_2+k_3$ and
\begin{eqnarray}
\hspace{-3mm}C_1 &=& Re[(\alpha^S_k-\beta^S_k)^3 ({{\alpha^S_k}^*}^{3}+{{\beta^S_k}^*}^{3})]\nonumber\\&\simeq& -\frac32 \cos2\phis\sin^2\phis e^{4\chis},\nonumber\\
C_2&=& Im[({\alpha^S_k}^*-{\beta^S_k}^*)^3 ({\alpha^S_k}^3-{\beta^S_k}^3)]\nonumber\\&\simeq& -\frac32 \sin2\phis\sin^2\phis e^{4\chis},\nonumber\\
C_3 &=& Re[(\beta^S_k-\alpha^S_k)^3 ({\alpha^S_k}^*{{\beta^S_k}^*}^2+{{\alpha^S_k}^*}^2{\beta^S_k}^*)]\nonumber\\&\simeq& -\frac12 (\cos2\phis+2)\sin^2\phis e^{4\chis},\nonumber\\
C_4 &=& Im[(\alpha^S_k-\beta^S_k)^3 ({{\alpha^S_k}^*}^2{\beta^S_k}^*-{\alpha^S_k}^*{{\beta^S_k}^*}^2)]\simeq -\frac23 C_2,\nonumber%
 \end{eqnarray}
In the second equalities, we have given the leading contribution for  \emph{typical or large} ${\chis}\gtrsim 1$. Recalling that
$H^4/\epsilon\propto {\cal P}_S^2\epsilon \gamma_{{}_S}^{-2}\propto {\cal P}_S^2\epsilon e^{-4\chis}/\sin^4\phis$ and $f_{\rm NL}\simeq {\cal B}/{\cal P}_S^2$ \cite{Maldacena:2002vr, Ashoorioon:2010xg} 
the powers of $e^{\chis}$ cancel out in the leading contribution. Since we are interested in cases where $\phis\simeq \pi/2$ we do not get any enhancement due to the $\sin\phis$ factors in the denominator of the expression for $f_{\rm NL}$.

For excited states two types of enhancement can occur: (i) {\it ``flattened configurations''} \cite{Chen:2006nt} (and subsequent work
\cite{Holman:2007na, folded-NG, Ashoorioon:2010xg}) in which $k_1+k_2\simeq k_3$ and two of the vectors are almost collinear. The enhancement for this configuration is lost for  slow-roll inflation after taking into account the effect of two dimensional projection of the bispectrum to the CMB surface \cite{Holman:2007na}. (ii) {\it ``local configuration''} \cite{Agullo:2010ws, squeezed-NG} in which $k_1\ll k_2\approx k_3$. In this case there is a possibility of enhancement from the slow-roll results by a factor of $k_2/k_1$ \cite{Agullo:2010ws}. We take $k_2$ to be the smallest wavelength probed by Planck, $\ell\simeq 2500$ and $k_1$ the largest scale at which the cosmic variance is negligible, $\ell\simeq 10$. For $\chis\gtrsim 1$ and $\varphi_S\gtrsim \pi/10$, one obtains an amount of  non-Gaussianity which is below the $2\sigma$ limit set by  Planck on local $f_{\rm NL}$. For example, if $M$ is taken
to be close to its maximal value, $M\simeq 20 H$, {\it i.e.} $\chis\gtrsim 1$ and assume that $\phis=\frac{\pi}{2}$, one obtains
\begin{equation}\label{flocal}
f_{\rm NL}^{\rm local}\simeq 0.43
\end{equation}
which is well within $1\sigma$ viable region of the Planck results for local non-Gaussianity. For the same values of $\chi_S\gtrsim 1$, only when $\phis \lesssim\frac{\pi}{10}$, the local non-Gaussianity goes beyond the $2\sigma$ limit on the local $f_{\rm NL}$. However, this range of $\phis$ is already ruled out from the backreaction bound, eq.\eqref{M/H}.

\section{Concluding remarks }

{High energy scale models of inflation, like those with concave monomial potentials, are in tension with the Planck data due to the large tensor over scalar ratio they predict.  In this paper, we showed that there exist regions of parameter space in the initial conditions for scalar and tensor perturbations where tensor-to-scalar ratio is suppressed with respect with to the corresponding Bunch-Davies value. In fact the backreaction constraint along with the COBE normalization for the amplitude of density perturbations puts an upper bound on how large the scale of new physics could be with respect to the inflationary Hubble parameter. For large deviations from the Bunch-Davies vacuum, the scale of new physics $M$, which is implemented through a crude cut-off model, could at most be around $\sim 20 H$. In this region of parameter space, the effect on tensor over scalar ratio is suppressive regardless of the details of the initial condition for tensor fluctuation. In such region, the amount of non-Gaussianity in the local configuration is compatible with the Planck data.}

Simple chaotic models, in particular $m^2\phi^2$ model, were/are of interesting not only endowed with simplicity and beauty but also through predicting a large amplitude for the tensor modes which could be in reach of current proposals for CMB B-mode polarization searches. Nonetheless, recent Planck data \cite{Planck-data} put these models under severe constraints to the extent that some considered the inflationary paradigm to be in trouble \cite{Ijjas:2013vea}. Excited initial state for tensor and scalar fluctuations is viable proposal, because  through which one can reconcile these models with the Planck data for $n_s$ vs. $r$ diagram, while respecting the bounds on non-Gaussianity.

\section*{Acknowledgements}

A.A. and K.D. (in part) are supported by the Lancaster-Manchester-Sheffield Consortium for Fundamental Physics under STFC grant ST/J000418/1.
GS is supported in part by DOE grant DE-FG-02-95ER40896. M.M.Sh.-J. would like to thank the Izmir Institute of Technology (IZTECH) where a part of this work was carried out.

\bibliographystyle{apsrev}

\end{document}